\documentclass[9pt,twocolumn,twoside]{pnas-new}
\templatetype{pnasinvited} % Choose template 
\title{Bringing Information Centric Networking to Challenged Environments\\
{\Large An overview of the Second Workshop on Future Internet Architecture for Developing Regions}}

\author[a,1]{Andrés Arcia-Moret}
\author[b,1]{Ioannis Psaras} 
\author[a,1]{Jon Crowcroft}

\affil[a]{University of Cambridge, UK}
\affil[b]{University College London, UK}

\correspondingauthor{\textsuperscript{1}E-mail: andres.arcia$@$cl.cam.ac.uk, i.psaras$@$ucl.ac.uk, jon.crowcroft$@$cl.cam.ac.uk}

\authorcontributions{}
\authordeclaration{}
\equalauthors{} 


\begin{abstract}
The 2nd workshop on Future Internet Architecture for Developing Regions (FI4D) took place on January 8th, 2017. The workshop was hosted by the 14th Annual IEEE Consumer Communications $\&$ Networking Conference in Las Vegas, USA. This report summarizes the motivation, discussions and perspectives held during the half-day event.  
\end{abstract}

\dates{This manuscript was compiled on \today}
\doi{$2^{nd}$ FI4D website: \url{https://goo.gl/4QHsnh}}

\begin{document}

% Optional adjustment to line up main text (after abstract) of first page with line numbers, when using both lineno and twocolumn options.
% You should only change this length when you've finalised the article contents.
\verticaladjustment{-2pt}

\maketitle
\thispagestyle{firststyle}
\ifthenelse{\boolean{shortarticle}}{\ifthenelse{\boolean{singlecolumn}}{\abscontentformatted}{\abscontent}}{}

\noindent \lettrine[findent=2pt]{\fbox{\textbf{T}}}{ }he Future Internet Architecture for Developing regions (FI4D) has reached its second edition! We have been honoured to have Dr Ioannis Psaras, EPSRC Fellow and Lecturer from University College London as the keynote speaker, four papers presented and an attendance of 20 people. Towards the end, we had an open forum discussing further questions about presented papers, emphasising on the need for integration of ICN and DTN – not only an ongoing research topic for the H2020 RIFE, H2020 UMOBILE and EPSRC INSP projects, but also a subject of research in the previous FP7 GreenICN project (2013-2016). We hope FI4D is going to get better this year, as we expect to bring it to Europe for the first time.

Before further elaboration on what happened at FI4D, we would like to thank the Internet Society, especially to Jane Coffin who has been very supportive of this initiative.

FI4D is about bringing together academic participants working on innovative solutions for bridging the gap in Internet coverage for everyone (especially in remote regions) using cutting-edge technologies such as ICN, SDN, NFV within modern motivational frameworks such as 5G networks. FI4D is an initiative originally covering the need for synergy between RIFE (\url{http://rife-project.eu}) and UMOBILE (\url{http://www.umobile-project.eu}); these two projects have the common objective of fostering Internet connectivity in developing regions. In \textbf{RIFE} we are developing a solution for bringing Internet connection to rural and remote areas at a low-cost, whereas in \textbf{UMOBILE} we are developing a solution to bring the Internet in disaster scenarios; both of them using an Information Centric Networking approach.

The necessary synergy between the two projects acts as the innovation driver in FI4D. At the University of Cambridge, we are bringing together RIFE and UMOBILE to promote solutions based on two prevalent ICN views: PURSUIT \cite{trossen} and NDN \cite{zhang}. In RIFE we count on PURSUIT/Blackadder as the backbone of our implementation for bringing fast and low-cost Internet to developing regions. In UMOBILE we count on NDN to provide a completely decentralized solution for disaster scenarios based on ICN. Reasons for choosing different implementations are straightforward when we see that to foster Internet access in remote areas, we mostly need a mid-size network, connecting small to medium communities (up to hundreds of people) and relying on local fast caching with an appropriate dissemination strategy for specific content (e.g., multicast for video on demand). On the other hand, disaster scenarios bring a different requirement for ICN, namely the long or permanent disconnected island of the network. In this regard, CCN can fully distribute content having the bits and pieces of the information spread out around the network. Having no single point of failure allows the spawning of fully operational sub-networks, maximising the possibilities of resilience in disaster scenarios. 

\section*{Alternative Network Deployments}

Based on the recently published RFC 7962 \cite{saldana}, in FI4D we believe that Alternative Network Deployments (ANDs) are not only the way to connect the next billion people but also a relevant case of study (and an essential building block) for disaster scenarios. As we expect, ANDs will also be a pervasive vision for infrastructure deployment in critical regions where flooding, earthquakes, and other natural disasters affect people the most. As ANDs evangelists trumpet: ``... we are willing to obtain the attention of communities seeking to create and manage computer networks for the people by the people'' \cite{arcia2}.

As stated in \cite{arcia2} Alternative Network deployments face three types of challenges: ``\textbf{geographic}, motivated by the need to connect rural and remote areas; \textbf{technological}, given the need for a common set of technologies that enable a better utilization of scarce resources; and \textbf{socioeconomic}, based on the need to study affordability models for disconnected people.''

These challenges have motivated in the last decade the so-called ANDs that are being perceived nowadays as Internet access models, not only intended to provide Internet connectivity, but more importantly to understand and study different topological, architectural, governance, and business models different from the so-called ``mainstream'' ones. Distinct from these models, ANDs do not expect users to pay regular and unaffordable fees to be connected and make use of the Internet.
As AND is quite a recent concept and approach, we believe that they can also be unusually defined by what they are not! They guarantee openness and privacy by not being conceived top-down and not being controlled by a central authority (who could eventually intervene communication). Moreover, they have no infrastructure with substantial investment. However, these restrictions do not impose a limitation in scalability, as GUIFI.net\footnote{\url{https://guifi.net/en}}  (in Catalunya, Spain), the largest Alternative Network in the world, can account for it. 

GUIFI.net being a partner of RIFE is the best example of what an Alternative Network can expect to be if the correct topology, motivation and governance models meet in one place. Nowadays, and after 10+ years of life, GUIFI has more than 30,000 active nodes, more than 60,000 Km of wireless links and is currently generating traffic of 5 Gbps. At the socio-economic scale, GUIFI has incubated more than 25 SMEs creating hundreds of direct jobs, and even more indirect jobs. Also, more importantly, it has promoted a self-sustainable model counting on hundreds of volunteers and public administrators willing to commit to working for the community.

Although GUIFI.net is hosted in a country in which the digital divide is almost non-existent, it is a guiding light to develop self-sustainable models that could help in the bridging of the digital divide to developing regions; thereby increasing the availability and affordability of the network infrastructure. Moreover, we expect ANDs helping with the digital illiteracy, promoting changes in the regulatory framework for the masses and popularising content and services - what, by the way, is a suitable approach to hosting ICN networks.

The inclusion of rural zones into the mainstream networks is one of the approaches being currently explored in RIFE. We are investigating the growth and conformation patterns of ANDs, as we see that they are most likely to be deployed around urban areas, which may act as centres of gravity. We further believe that Community Networks (CNs) are the most suitable type of AN to follow this pattern. 

As discussed in the RFC 7962, successful CNs (e.g., GUIFI) have been conceived so that participants can have the freedom to use the network for any purpose while at the same time respecting their fellow networkers. CNs count on a self-contained set of rights on the whole infrastructure guaranteeing a stable governance. In a nutshell, these rights include (see \cite{saldana} for more information): the right of understanding the network and its design principles, the right of offering services on people's terms, and the right to be part of the network so as to extend the service under the same conditions. In this context, the Ostrom principles have provided further guidance on the governance of the common pool of resources (CPR) \cite{baig}, which we consider one of the main challenges in CNs as governance structure can depend on the culture.

All in all, we should keep in mind that Alternative Networks are mainly intended to guarantee a distributed administration to serve rural underserved areas and disconnected network islands, most likely through wireless technologies in unlicensed portions of the spectrum.

\section*{The RIFE Approach to Efficient Information Dissemination in Alternative Networks}

The Architecture for an Internet for Everybody (RIFE) is a Horizon 2020 project working along three main axes. 

\textbf{Socio-economic}. We are investigating new ways of directly benefiting the community providing affordable technology solutions. We have developed a value network configuration approach by analysing and putting together all the necessary basic pieces given different possible technological approaches to initiate a RIFE-like network in developing regions \cite{benseny}. We expect to evaluate this approach in the Camp de Tarragona community in Spain, which has been recently chosen to be the host of our field trial in RIFE, benefiting more than 10.000 people from one city and several towns. 

\textbf{Affordability}. In RIFE we are integrating several technologies aimed to facilitate affordable access, by minimising deployment costs and opportunistically maximising the use of networking resources. This is why we are integrating into RIFE the SCAMPI Delay Tolerant Network solution \cite{pitkanen}, elaborating new low-cost network access plans through satellite (with AVANTI), and leveraging on Alternative Network Deployments (aka Community Networks) not only to lower costs but also to provide new governance structures that empower communities with technology and knowledge.

\textbf{Scalability}. As we are leveraging our solution in PURSUIT/Blackadder implementation, we need to address the inherent limitations of routing and governance. We are extending the link-addressing strategy to increase the number of addressable links within a PURSUIT network. This extension allows us, in turn, to realise a different governance model, allowing different stakeholders to have independence in their routing strategy. However, in the end, conflicting interests can be reconciled through a distributed approach of topology management \cite{arcia1}.

In RIFE we see unique opportunities to have open testbeds relying on ANDs to tests not only large systems such as PURSUIT/ICN but also simpler protocols and networking approaches (e.g., new congestion control protocols). These testbeds are easily extensible since they can be implemented using overlay networks. Finally, we expect to see shortly expedited and efficient governance models fostered by ANDs having network managers that, contrary to mainstream network approaches (in governance), facilitate the deployment of local services and extensions of the Alternative (and Community) Networks.

\subsection*{Challenges on Inter-domain Forwarding and Privacy}

PURSUIT ICN focuses on \textit{dissemination strategies} as a key concept for tackling the efficient dissemination of information - thus RIFE leverages on this concept to further increase scalability and address privacy requirements. In RIFE, we are designing the \textbf{Zoning} and \textbf{Inter-Zoning} dissemination strategies that enable not only having larger networks under a single domain, but also allowing independent communities and operators to co-exist in different domains; hopefully, also promoting self-sustainability as ANDs can enable competition among providers through the generation of value. RIFE Inter-Zoning dissemination strategy will allow independent organizations and operators to provide the backhaul service while guaranteeing privacy (e.g., concealing their topology) through the introduction of a distributed topology management function and adaptable link-labelling strategies.

\section*{Information Centricity for Local Community Networks and Disaster Management}

Soon after the establishment of Information-Centric Networks as an active topic of research, the community realised that ICN could bring significant benefits in mobile communications. In 2013, leading partners in the ICN area from Europe and Japan came together to form a consortium for the GreenICN project (Architecture and Applications for Green Information Centric Networking). The collaborative EU FP7/JP NICT project focused on two exemplary use-cases: i) disaster management and ii) mobile video delivery. The Great East Japan Earthquake in 2011 has forced the Japanese government to set new targets for communication networks.  Among others, there was a strict requirement to reduce the network’s power consumption by 20\% in normal days and by 40\% after disasters. 

The consortium produced several solutions on this direction, some of which were presented or discussed during the FI4D workshop \cite{psaras1,ohsugi,tagami,seedorf}. Intrinsically, these solutions perfectly fit ANDs, and were the spin during the FI4D workshop. Although the solutions presented and discussed were based largely on the NDN paradigm, their validity is not limited to NDN’s operational principles. For instance, ``Name-Based Replication'' (NREP) is based on the concept of ``Information-Centric Connectivity'' (ICCON) \cite{katsaros}. According to this concept, mobile users connect and pair with other users or Access Points (APs) that are likely to have the requested content. Results, indeed, show great performance improvement compared to content-agnostic connectivity. Exploiting this notion in the mobile domain results in replication of content among mobile devices (e.g., smartphones) based on the name (and properties or even cryptographic hash) of the content that is being disseminated. Apart from cache-hit rate and successful dissemination benefits, NREP results in reduced energy consumption, as it replicates (or connects according to ICCON) only content that is of interest.

\subsection*{The UMOBILE Approach for Disaster Management}

Universal, mobile-centric and opportunistic communications architecture (UMOBILE) is a Horizon 2020 project developing an architecture that integrates the principles of Delay Tolerant Networking (DTN) and Information Centric Networking (ICN) in a common framework.

UMOBILE utilizes the benefits of both ICN and DTN to enable resource exploitation at minimal bandwidth, opportunistic access to information and more localized access to information through novel caching strategies.

UMOBILE focuses on assisting users in getting access to the content they want or content that may be of shared interest to their trust circles. By relying on an instance of the UMOBILE architecture, users can share information or computation directly with other peers without relying on infrastructure or expensive connectivity services.

Along those lines, in UMOBILE, we have developed (among others) a radical approach to edge-computing. According to our “Keyword-Based Mobile Application Sharing” \cite{psaras2} framework (KEBAPP) users share their mobile smartphone apps with users in the vicinity. When a user requires some service (e.g., recommendation on restaurants, or the weather forecast), they will typically have to connect to the distant cloud. More often than not, however, such information is available on nearby devices, especially in urban, densely populated areas. Our novel ICN-based request pattern, which allows for free keywords to be included in the message can resolve requests to the right application (in nearby devices) and invoke computation to serve the request. We find that response times can be lower than the ones experienced over the cellular network (in developed countries). We have also argued that in the presence of limited funding, which is certainly the case for ANDs and community networks (in either developing or developed countries), the KEBAPP approach can serve as an alternative cloud for local information of common interest.

\section*{FI4D paper session}

During the paper session, we had two main topics of discussion: naming and congestion control for ICN networks; both of them timely on the current interests of ICN community, and on the interests of RIFE and UMOBILE projects. 

\subsection*{Naming in ICN Networks}

As the famous quote attributed to Phil Karlton goes: “The only real difficulties in computer science are cache invalidation and naming things”. This is indeed very close to the interests of ICN theory, making information as the centre of networking thus naming comes as a first player. During the first session, we had the discussion on the scalability of naming and the design and validation of a system for mobility handling with wildcard names. 

Urs Schnurrenberger \cite{urs} presented ``Comparing Apples to Apples in ICN''. Urs discussed the need to have the right models describing ICN names by length and structure. For that, he presented an analysis of data sets ranging in the billion of entries to characterise ICN naming in the wild (using URL type of names). Interestingly enough, the mean and median number of names to characterise content in such datasets is around 5, with an average length in a number of characters of about 90. He also managed to generate fairly precise mathematical models abstracting the huge amount of processed data. Urs showed an interesting future application for Alternative Networks and his findings encourage to use commodity and low-cost/low-capacity device being capable of hosting regular content names (for large data representations) as they seem to fit small memory spaces. 

Daiki Ito \cite{ito} presented ``Virtual Storage and Area Limited Data Delivery over Named Data Networking''. Daiki introduced a system capable of transparently handling mobility in ICN networks. He emphasised on a hierarchical naming structure and to provide some particular keywords giving a transparent vision of mobility. For example, by referring to a certain information item as ``.../here/1.txt'' the system allows a user to obtain an updated and local copy of the requested information item. The system was evaluated in a vehicular network scenario, but one can easily envision this same system providing a mobility service for Community Networks geographically separated and potentially sharing content.

\subsection*{Congestion Control in Future Networks}

Congestion control is a hot topic nowadays in ICN networks. As the information is likely to be spread around the network in NDN, and given the heterogeneous characteristics of Alternative Network scenarios, hop-by-hop communication solutions are gaining popularity. In this respect, we had the presentation of two approaches to deal with high capacity networks in 5G scenarios and to deal with hop-by-hop congestion control in NDN networks. 

Ivan Petrov \cite{petrov} presented ``Advanced 5G-TCP: Transport Protocol for 5G Mobile Networks''. Ivan introduced a novel approach to deal with high bandwidth-delay product networks in 5G contexts. Departing from the well-known High-Speed TCP, authors propose a transport protocol manager strategically located on the network and able to instruct different portions of the network (e.g., at the edge of femtocells) to use the most convenient congestion control function depending on the BDP. So, higher BDPs will be controlled by more aggressive growth functions obtained from a pool of parabolic pre-calculated functions based on well-known network characteristics. Ivan presented an interesting exercise on best adapting a congestion control function to maximise the throughput in femtocells when undesired losses occur, thereby compensating the unnecessary overcut of the congestion window. 

Finally, Takahiko Kato \cite{kato} presented ``Congestion Control Avoiding Excessive Rate Reduction in Name Data Network''. Takahiko introduced a hop by hop congestion control and presented their solution to unnecessary windows cut by proposing a special treatment of excessive NACK packets, mistakenly reducing the sender's throughput. By a simple state machine that identifies the first loss and waiting until a reordering episode finishes, authors recreate the Tahoe to Reno-TCP evolution story in ICN networks using a hop-by-hop approach. We hope that many interesting experiments and extensions can follow upon this work.

\section*{Open Forum Discussion}

In the end, we had an open discussion about the all the topics presented during the half-day session. Most of the insights  have been summarized throughout the document. However, there were open questions about naming and ICN which generated enthusiasm given RIFE and UMOBILE current state of development. In both projects, there is a common interest in having a unified naming interface integrating the core network and the disconnected parts of the network (i.e., as in DTN scenarios). On the other hand, with respect to congestion control, there was a discussion about rate based congestion control for which Ioannis vision motivated the talk \cite{psaras3}.

\acknow{We would like to thank to all members of the Technical Program Committee, and to Marco Zennaro and Adisorn Lertsinsrubtavee co-chairs of the workshop. Special thanks to Jane Coffin and the Internet Society for supporting the FI4D initiative. Thanks also to Roger Baig who always provides insights about Community Networks vision.}

\showacknow % Display the acknowledgements section

% \pnasbreak splits and balances the columns before the references.
% Uncomment \pnasbreak to view the references in the PNAS-style
% If you see unexpected formatting errors, try commenting out this line
% as it can run into problems with floats and footnotes on the final page.
% \pnasbreak

% Bibliography
%\bibliography{pnas-sample}

\end{document}